\documentstyle[12pt,epsf]{article}
\def\fun#1#2{\lower3.6pt\vbox{\baselineskip0pt\lineskip.9pt
        \ialign{$\mathsurround=0pt#1\hfill##\hfil$\crcr#2\crcr\sim\crcr}}}

\begin{document}

\title{\vskip-2.5truecm{\hfill \baselineskip 14pt {{
\small  \\       \hfill SISSA--170/96/EP \\ \hfill FTUV/96-81 \\
\hfill IFIC/96-90 \\
\hfill November 1996}}\vskip .1truecm}
 {\bf Baryogenesis from mixed particle decays}}

\vspace{2cm}

\author{Laura Covi$^{1}$ and Esteban
Roulet$^{2}$
 \\  \  \\
 $^{1}${\it International School for Advanced Studies, 
SISSA-ISAS},\\ {\it Via Beirut 2/4, I-34014, Trieste, Italy} \\ 
 $^{2}${\it Depto. de F\'\i sica Te\'orica, Universidad de Valencia,}\\
{\it E-46100, Burjasot, Valencia, Spain}
}

\date{}
\maketitle
\vfill

\begin{abstract}
\baselineskip 20pt

We consider the $CP$ violating asymmetries produced in the decay of
heavy particles, studying the effects of heavy particle mixing for
arbitrary mass splittings. A considerable enhancement of
the asymmetries is achieved when the masses of the mixed states are
comparable, and the enhancement is maximal for mass splittings of
the order of the widths of the decaying particles. We apply the
results to the particular case of heavy scalar neutrino decays
relevant for leptogenesis scenarios.

\end{abstract}
\vfill
\thispagestyle{empty}

\newpage
\pagestyle{plain}
\setcounter{page}{1}
\def\beq{\begin{equation}}
\def\eeq{\end{equation}}
\def\beqa{\begin{eqnarray}}
\def\eeqa{\end{eqnarray}}
\baselineskip 24pt
\parskip 8pt
 
The violation of $CP$ is one of the crucial ingredients identified by
Sakharov \cite{sa67} as necessary conditions for a dynamical generation of the
baryon asymmetry of the Universe. Since the known $CP$ violation in the
standard model is probably too small to be helpful in generating the
observed baryon asymmetry, the existence of new $CP$ violating
interactions in most of the extensions of the standard model is
particularly welcome from this point of view.
In the usual scenarios of baryon (or lepton) number generation by the
out of equilibrium $B$ (or $L$) violating decays of heavy particles, the
$CP$ violation arises in general from the interference of the tree level
and one loop decay amplitudes, which allow the phases in the complex 
couplings  involved to show up in the partial decay rate asymmetries.
The one loop contribution which is usually taken into account is the
so called `vertex' one, in which two light particles produced in the
decay of a heavy one exchange another heavy particle to produce  the
required final state (Fig.~1b). However, many of the scenarios studied
involve (and even require\footnote{The existence of just one heavy
triplet scalar in the minimal SU(5) theory was actually considered a
problem as regards the baryon number generation, since the $CP$
violation appeared only at three loops \cite{ha82a}.}) more than one
`flavour' of heavy particles, allowing then for a further possibility
to produce $CP$ violation.

This new possibility arises from the so called `wave--function'
contribution \cite{ig79,bo91,li93,fl95,co96}, 
in which a loop of light particles just mixes the
initial state $\Phi_a$ with another different heavy state $\Phi_b$,
which then decays to the final state as shown in Fig.~1c. This wave
contribution turns out to be comparable to the vertex one when the
heavy states have large mass splittings, and may be significantly
enhanced for nearly degenerate states.

The asymmetry in a global quantum number $N$
(for instance $B$ or $L$) produced in the decay of a pair made of 
particle $\Phi_a$
and its antiparticle $\bar \Phi_a$
is given by
\beq
\epsilon_a=\sum_f \epsilon_{fa},
\eeq
where the quantity of interest to us is the partial rate asymmetry per decay
into final state $f$ (of global charge $N_f$), given by
\beq
\epsilon_{fa}=N_f\left[\mbox{BR}(\Phi_a\to f)-\mbox{BR}(
\bar \Phi_a\to \bar f)\right],
\label{eqepsfa}
\eeq
where $\mbox{BR}(\Phi_a\to f)$ is just the branching ratio for the
decay of particle $\Phi_a$ into the final state $f$.
The wave function contribution to this quantity behaves, in the limit
of large mass splittings, as $\epsilon^w_{fa}\propto
(M_a^2-M_b^2)^{-1}$, due to the propagator of the intermediate state
$\Phi_b$, and hence is expected to be enhanced in the limit $M_b\to
M_a$. A general approach to study this quantity for arbitrary mass
splittings has been considered in ref.~\cite{li94}. It is our
purpose here to extend the formalism developed in ref.~\cite{li94}
 and apply the results to the study of 
leptogenesis scenarios, for which a computation of the $CP$ violation
for large mass splittings was recently obtained \cite{co96}. 

To be specific, we will consider the case in which the heavy decaying
particles are scalars, and ignore the vertex $CP$ violation effects,
which can be studied separately. The wave function mixing will have
the effect of inducing an absorptive part in the heavy particle
propagators, which will be responsible for the generation of the
asymmetry. The effect of the one--loop self--energy 
diagrams in the propagators
will be to modify the squared
scalar mass matrix as follows
\beq
M^{(0)2}_a\delta_{ab}\to H^2_{ab}=M^2_{ab}-i\Gamma^2_{ab},
\label{hq}
\eeq
where the renormalised mass matrix $M^2$ includes the dispersive part
of the loops while the matrix $\Gamma^2$ arises from the absorptive
part alone.
The matrices $M^2$ and $\Gamma^2$ are hermitian, but $H^2$ is not.
Hence, $H^2$ will be diagonalised in general by a non--unitary
transformation matrix $V$ \cite{li94}, i.e.
\beq
(VH^2V^{-1})_{ab}=\omega^2_a\delta_{ab}.
\label{hqd}
\eeq
This matrix $V$ will then transform the initial `flavour' states
$|\Phi_a\rangle$ into the `propagation' eigenstates\footnote{The 
appearance of $V^{-1}$ 
 in eq.~(\ref{hqd}) ensures that the kinetic term remains canonical, but
the fact that 
$V^{-1}\neq V^\dagger$ implies that the propagation eigenstates are
not orthonormal.} 
$|\Phi_c'\rangle$, i.e. 
\beq
|\Phi_c'\rangle=V^{-1}_{ac}|\Phi_a\rangle.
\eeq
Similarly, for the antiparticle states $|\bar\Phi_a\rangle$, one will
have 
\beq
|\bar\Phi_c'\rangle=V_{ca}|\bar\Phi_a\rangle.
\eeq
These propagation eigenstates are the ones that will evolve as
\beq
|\Phi_c'(t)\rangle=\mbox{e}^{-i\omega_c t}|\Phi_c'(0)\rangle.
\eeq
Considering then the transition amplitude from the state
$|\Phi_a\rangle$  to a
final state $|f\rangle$, we have
\beq
T_{fa}=\langle f|H_{int}|\Phi_a\rangle,
\eeq
where $H_{int}$ describes the interactions of $\Phi_a$ with the final
state particles. From the superposition principle and using
Eqs.~(\ref{hq},\ref{hqd}), one has
\beqa
T_{fa}(t)=&\sum_{b,c}T_{fb}V^{-1}_{bc}V_{ca}\mbox{e}^{-i\omega_ct}\cr
\bar T_{fa}(t)=&\sum_{b,c}T^*_{fb}V_{cb}V^{-1}_{ac}\mbox{e}^{-i\omega_ct}.
\label{eqttbar}
\eeqa
The differential partial decay rate asymmetries arising from particle
mixing will be proportional to the quantities
\beq
\Delta_{fa}(t)\equiv |T_{fa}(t)|^2-|\bar T_{fa}(t)|^2.
\label{eqdfa}
\eeq
It is interesting to notice that, in the limit of degenerate
propagation eigenstates, i.e. $\omega_c=\omega$, these asymmetries 
vanish, as can be seen from eqs.~(\ref{eqttbar},\ref{eqdfa}).

To continue we will concentrate in the case of mixing between just two
particles, for which the matrix $V$ can be parameterised in terms of
two complex mixing angles, $\theta$ and $\phi$, as follows
\beq
V=\pmatrix{\mbox{cos}\theta & -\mbox{sin}\theta\mbox{e}^{i\phi}\cr 
\mbox{sin}\theta\mbox{e}^{-i\phi} & \mbox{cos}\theta}.
\eeq
Replacing this in Eq.~(\ref{hqd}) it is easy to obtain
\beq
\mbox{e}^{2i\phi}={H^2_{12}\over H^2_{21}}\ \ ;\ \
(\mbox{tg}2\theta)^2= {4 H^2_{12}H^2_{21}\over (H_{11}^2-H_{22}^2)^2},
\eeq
where we recall that $H_{21}^2=M^{2*}_{12}-i\Gamma^{2*}_{12}$.
The eigenvalues of $H^2$ are then
\beq
\omega^2_{1,2}={1\over 2}\left\{ H^2_{11}+H^2_{22}\pm
\sqrt{(H^2_{11}-H^2_{22})^2 +4H^2_{12} H^2_{21}}\right\} .
\eeq
After an explicit computation we then get
\beq
\Delta_{fa}(t)=2 \mbox{Re}\left\{T_{f1}T^*_{f2}\left[U_{1a}U^*_{2a}-
U_{a2}U^*_{a1}\right]\right\} +
|T_{fb}|^2\left\{|U_{ba}|^2-|U_{ab}|^2\right\} , 
\eeq
where  we have defined
\beq
U_{ab}\equiv V^{-1}_{ac}W_cV_{cb},
\eeq
with     
\beq
W_c\equiv \mbox{e}^{-i\omega_ct}.
\eeq

In the case in which the initial state under consideration is an
eigenstate of the (renormalised) mass matrix, i.e. for $M^2_{ab}= M^2_a 
\delta_{ab}$, these expressions simplify considerably, and we have
\beq
\Delta_{f1}(t)=4{\mbox{Im}\left\{ T_{f1}T^*_{f2}\Gamma^2_{12}\right\}
\over |\omega_1^2-\omega_2^2|^2}\mbox{Re}\left\{
(\omega_1^2-\omega_2^2) (W^*_2-W^*_1)(\mbox{cos}^2\theta
W_1+\mbox{sin}^2\theta W_2)\right\}, 
\eeq
and a similar result holds for $\Delta_{f2}$ with the substitution
$W_2\leftrightarrow W_1$.

We will then compute in detail the integrated rate asymmetry in this
case. 
The branching ratios entering in eq.~(\ref{eqepsfa}) are just
\beq
\mbox{BR}(\Phi_a\to f)=\int
d\Omega_a\int_0^\infty dt\ |T_{fa}(t)|^2,
\eeq
with $d\Omega_a$ the
phase space element of particle $\Phi_a$. We have then
\beq
\epsilon^w_{fa}=N_f\Omega_a\int_0^\infty dt \Delta_{fa}(t),
\eeq
where $\Omega_a=M_a/16 \pi$ in our case of two body scalar decay.

Integrating over time we find
\beq
\epsilon^w_{fa}=2N_f\Omega_a \mbox{Im}\left\{
T_{f1}T^*_{f2}\Gamma^2_{12}\right\} F_a,
\label{eqepsw}
\eeq
with
$$
F_1={1\over |\omega_1^2-\omega_2^2|^2}\left\{
\mbox{Re}\{\omega_1^2-\omega_2^2\}
\left[{1\over\gamma_2}-{1\over \gamma_1}
\right] -(M^2_1-M^2_2)
\left[{1\over \gamma_1}+{1\over \gamma_2}- {\gamma_1+\gamma_2\over 
|\omega_1^*-\omega_2|^2}\right]\right. 
$$
\beq
+\left.
2\ \mbox{Im}\{ \omega_1^2-\omega_2^2\} {(m_1-m_2)\over
|\omega_1^*-\omega_2|^2}\right\},
\label{eqf1}
\eeq
where we defined $\omega_a\equiv m_a-i\gamma_a/2$. The result for
$F_2$ is similar with the replacements
$\gamma_1\leftrightarrow \gamma_2$ and $m_1\leftrightarrow m_2$ in
the expression for $F_1$. 

In the case $M^2_2\gg M^2_1\gg |\Gamma^2_{ab}|$, one has
\beq
F_a\to {2\over \gamma_a (M^2_2-M^2_1)},
\eeq
where in this limit
 $\gamma_a \simeq \Gamma^2_{aa}/M_a$ becomes just the total width of
particle $\Phi_a$. 
Hence,  the results obtained in the literature in the limit of large
mass splittings can be recovered.

For decreasing mass splittings, the function $F_1$ reaches a maximum,
which for $|\Gamma^2_{12}|\ll |\Gamma^2_{22}-\Gamma^2_{11}|$ takes place
for
\beq
M^2_2-M^2_1\simeq \Gamma^2_{11}+\Gamma^2_{22}.
\label{eqdmmax}
\eeq
The value of $F_1$ at the
maximum is $F_1\simeq
M_1/[\Gamma^2_{11}(\Gamma^2_{11}+\Gamma^2_{22})]$.

On the other hand, for $|\Gamma^2_{11}-\Gamma^2_{22}|\gg 
|M^2_1-M^2_2|,|\Gamma^2_{12}|$, i.e. in the limit of small mass 
splittings and small mixing,   we get
\beq
F_1\simeq {(M^2_2-M^2_1)\over |\Gamma^2_{11}-\Gamma^2_{22}|^2}
{2M_1\over \Gamma^2_{11}}\left\{1-{4\Gamma^2_{11}\Gamma^2_{22}\over
(\Gamma^2_{11}+\Gamma^2_{22})^2}\right\} -{1\over M_1
(\Gamma^2_{11}+\Gamma^2_{22})}.
\label{eqf1b}
\eeq
This result coincides with the one in
ref.~\cite{li94} only in the limit $\Gamma^2_{11}\ll\Gamma^2_{22}$ (or
$\Gamma^2_{11}\gg\Gamma^2_{22}$) and if we neglect the last term, which
although small is non--vanishing and survives in the limit $M^2_2\to
M^2_1$ (in which the propagation eigenstates are not degenerate due
to $\Gamma^2_{11}\neq \Gamma^2_{22}$).

Another simple example is for the case $\Gamma^2_{11}=\Gamma^2_{22}$,
for which we have
\beq
\omega^2_2-\omega^2_1=\sqrt{(M^2_2-M^2_1)^2-4|\Gamma^2_{12}|^2}
\eeq
We see that here for $|M^2_2-M^2_1|=2|\Gamma^2_{12}|$ the two
propagation eigenstates become degenerate, and in fact for this mass
splitting $\epsilon^w_{fa}=0$, but it will be non--vanishing otherwise.
In particular, in the limit\footnote{The `degenerate' situation
$\Gamma^2_{11}=\Gamma^2_{22}$ and $M^2_1=M^2_2$ actually is
present in well known cases such as $K^0\bar K^0$ or $B^0\bar
B^0$ mixing, where those constraints are imposed by $CPT$ relations,
and for which the integrated $CP$ violating asymmetries are
non--vanishing \cite{du86}.}
 $M^2_2\to M^2_1$ one has $F_1\simeq
-1/(2M_1\Gamma^2_{11})$.  

Let us also emphasise that a crucial ingredient in all this
computation is the proper specification of the initial state. The
asymmetry of course depends on the starting basis for $\Phi_a$
considered, and hence on the process which produces the initial state,
so that ignoring the mixing at production would lead to incorrect
results. For instance, in the case in which $M^2\propto {\bf 1}$ (or more
generally whenever the matrices $M^2$ and $\Gamma^2$ commute), it is
possible to change basis with a unitary transformation to make 
$M^2$ and $\Gamma^2$ both diagonal. Hence, the asymmetry computed
in this new basis will vanish, since \linebreak
$\Gamma^2_{12}=0$ now. However, these
new states may not be the quantum states generated in the production
process, and therefore the new basis may not be the appropriate one to
compute the resulting asymmetry.

Let us now consider the particular example of lepton
number generation in the out of equilibrium decay of heavy scalar
neutrinos, i.e. the supersymmetric version of the Fukugita and
Yanagida scenario \cite{fu86}. In these type of models
\cite{lu92,va92,ca93,mu93,pl96},  the decay of the
electroweak singlet (s)neutrinos, with masses $M\gg \mbox{TeV}$,
produces a lepton asymmetry. This is then partially converted into
a baryon asymmetry \cite{kh88}
 by the effects of the anomalous $B+L$ violation in the
standard model \cite{ma83},  which is in equilibrium at
temperatures larger than the electroweak phase transition one ($\simeq
10^2$~GeV). The study of the $CP$ violation in these models,
considering both the
`vertex' part as well as the `wave' contribution in the limit of large
mass splittings, was carried out recently \cite{co96}. 
We now consider the effects
of mixing for arbitrary masses using the formalism introduced
above\footnote{There has been a recent attempt  to study the asymmetry
in heavy Majorana neutrino decays in the limit of small mass
splittings \cite{fl96}, but those results are however at variance with
 ours. For instance, we find a dependence on $H_{12}$ through
$H_{12}H_{21}$, and not through $(H_{12}+H_{21})^2$ as in
ref.~\cite{fl96}, although we expect similar results for neutrino and
sneutrino decays in the supersymmetric model considered here.}.

The Lagrangian for the scalar neutrinos is, in a basis in which the
mass matrix is diagonal,
\beq
{\cal L}=-\lambda_{ia}\epsilon_{\alpha\beta}\left\{M_a\tilde
N^*_a\tilde L^\alpha_iH^\beta+
\overline{(\tilde h^\beta)^c}P_L\ell_i^\alpha \tilde N_a
  \right\}+h.c.
\eeq
where $\ell_i^T=(\nu_i\ l^-_i)$ and $H^T=(H^+\ H^0)$ are the lepton and
Higgs doublets
($i=e,\mu,\tau$, and $\epsilon_{\alpha\beta}=-\epsilon_{\beta\alpha}$, with
$\epsilon_{12}=+1$).

Since we are interested in the possible implications of small mass
splittings, we will assume that the right handed neutrino masses
consist of two almost degenerate states, with the third one being much
heavier and hence effectively decoupled from the mixing mechanism.
 In this
case, the effects of the third scalar neutrino can be included
independently, and  the mixing effects can be studied with the two
flavour  formalism
discussed before. It is particularly
interesting that scenarios with this type of spectrum have been widely
considered in the literature \cite{ha82}, 
and can naturally arise in  $SO(10)$ models. 

We will assume that sneutrinos are produced out of equilibrium 
by a certain unspecified mechanism (e.g. if sneutrinos are 
inflaton decay products \cite{ca93} or the inflaton itself
\cite{mu93}), 
 and for simplicity consider that the
states produced initially correspond to one of the
eigenstates\footnote{If both $\tilde N_1$ and $\tilde N_2$ are
simultaneously, but incoherently, produced, one needs just to add the
asymmetries from both decays.}, say $\tilde
N_1$, of the mass matrix (so that $M^2_{12}=0$). Hence, the asymmetry
will  be
given by eqs.~(\ref{eqepsw},\ref{eqf1}), where in this model
 a direct computation of the absorptive part of the sneutrino
propagator leads to \cite{co96} 
\beq
\Gamma^2_{ab}={1\over 8\pi}\left[(\lambda^\dagger \lambda)_{ba}M_aM_b+ 
(\lambda^\dagger \lambda)_{ab}s\right],
\label{eqgab}
\eeq
where the square of the four momentum will just be $s\equiv p^2=M_1^2$ by the
on--shell condition. The first contribution to the r.h.s. of
eq.~(\ref{eqgab}) is given by the slepton and 
Higgs loop, while the second by the lepton and Higgsino one.

As final states, we need to consider two possibilities, i.e. 
$f= \tilde
L^\alpha_i H^\beta$ as well as $f= \bar\ell^\alpha_i \tilde
h^{c\beta}$. For the final state with sleptons, we have
$T_{fa}=-i \epsilon_{\alpha\beta}\lambda^*_{ai}M_a/\sqrt{s}$,  so that
only the second term in the r.h.s. of eq.~(\ref{eqgab}) contributes to
the total asymmetry, and we have
\beq
\sum_{i,\alpha,\beta} 
L_f\mbox{Im}\left\{T_{f1}T^*_{f2}\Gamma^2_{12}\right\}=
{M_1M_2\over
4\pi}\mbox{Im}\left\{(\lambda^\dagger\lambda)_{12}^2\right\}
=-{M_1M_2\over
4\pi}\mbox{Im}\left\{(\lambda^\dagger\lambda)_{21}^2\right\},
\eeq
where $L_f=+1$ is the lepton number of the final state.
On the other hand, for the decay $\tilde N_a\to \bar\ell^\alpha_i \tilde
h^{c\beta}$, one has $T_{fa}=-i \epsilon_{\alpha\beta}\lambda_{ai}$, and
only the first term in the r.h.s. of eq.~(\ref{eqgab}) contributes to
the total asymmetry. Since now $L_f=-1$, we end up with the same 
contribution as the one coming from  the slepton
channel.
(Notice that the asymmetry in a given final state channel results
from the mixing generated by a loop involving the particles
 of the other final state channel.)

So, summing the contributions from both final states we get
\beq
\epsilon^w_1=-{M_1^2M_2\over
16\pi^2}\mbox{Im}\left\{(\lambda^\dagger\lambda)_{21}^2 \right\}F_1,
\eeq
where $F_1$ is given in eq.~(\ref{eqf1}). 
If we use that
$\Gamma^2_{11}=(\lambda^\dagger\lambda)_{11}M_1^2/4\pi$, and the
asymptotic expressions discussed previously for 
$M^2_2\gg M^2_1\gg |\Gamma^2_{ab}|$, one can see that in this limit 
\beq
\epsilon^w_1=-{1\over 2\pi}{M_1M_2\over M^2_2-M^2_1}{\mbox{Im}\left\{
\left(\lambda^\dagger \lambda\right)_{21}^2\right\}\over 
\left(\lambda^\dagger \lambda\right)_{11}},
\label{epswa}
\eeq
which coincides with the expression obtained in ref.~\cite{co96}.

In figure 2 we plot 
 the total asymmetry $\epsilon^w_1$ for arbitrary
mass splittings, normalised
to the vertex contribution $\epsilon^v_1$ arising from the exchange 
of the second state $\tilde N_2$ \cite{co96}
\beq
\epsilon^v_1=-{1\over 4\pi}\ g\left({M_2^2\over
M_1^2}\right){\mbox{Im}\left\{
\left(\lambda^\dagger \lambda\right)_{21}^2\right\}\over 
\left(\lambda^\dagger \lambda\right)_{11}},
\eeq
where $g(x)\equiv\sqrt{x}\ \mbox{ln}[(1+x)/x]$.

Notice that $\epsilon^v_1$ contains actually the same
combination of Yukawas appearing in the wave function contribution
$\epsilon^w_1$ 
(i.e. the factor $\mbox{Im}\{ (\lambda^\dagger\lambda)_{21}^2\}$), so
that the $CP$ violating phase cancels in the ratio.

In fig.~2 we adopted for definiteness $\Gamma^2_{22}=M_1^2/10$, 
$\Gamma^2_{11}=|\Gamma^2_{12}|=\Gamma^2_{22}/10$, plotting the result
as a function of $x\equiv M_2/M_1$. In the limit of large mass
splittings, the wave contribution approaches twice the value of the
vertex one, as expected \cite{co96}. For decreasing
mass splittings ($x\to 1$), the enhancement in the wave contribution due
to the mixing of the states is apparent, and reaches a maximum value
$\simeq M_1^2/(\Gamma^2_{22}\mbox{ln}2)$ 
 for  $M^2_2-M^2_1\simeq  \Gamma^2_{22}$ (corresponding to
 $x\simeq 1.05$ in this
case) as discussed in eq.~(\ref{eqdmmax}).

The dotted line corresponds to the asymptotic expression for the wave
contribution in eq.~(\ref{epswa}), and
gives a reasonable approximation to the result for
$M^2_2-M^2_1>4\Gamma^2_{22}$.
For smaller values of $\Gamma^2_{22}/M^2_1$, the enhancement in the
wave contribution is
 larger (and  can be in principle of many orders of magnitude), and 
the asymptotic expression
is valid down to smaller values of $x$. 
On the other hand, for $M_2^2-M_1^2<\Gamma^2_{22}$, $\epsilon^w_1$
decreases significantly, and for $M_2\to M_1$ one has 
\beq
{\epsilon^w_1\over \epsilon^v_1}\to -{\Gamma^2_{11}\over
(\Gamma^2_{11}+\Gamma^2_{22})\ \mbox{ln}2},
\eeq
which is tiny. 
The results
are quite insensitive to the actual values of $\Gamma^2_{11}$ and
$|\Gamma^2_{12}|$, as long as this last stays much smaller than
$\Gamma^2_{22}$, so that the mixing angle $\theta$ is small. If 
$|\Gamma^2_{12}|\sim \Gamma^2_{22}$, the maximum enhancement is
somewhat smaller but the general behaviour remains similar.

It is important to keep in mind that the vertex and wave
contribution arising from the exchange of the heavier third scalar
neutrino $\tilde N_3$ may however be larger than the one coming from
the exchange of the second one $\tilde N_2$, even taking into account
the possible enhancements for small mass splittings, due to the
probably larger Yukawa couplings involved and the unknown size of the
$CP$ violating phases appearing in both channels (for three families,
there are actually three independent $CP$ violating phases entering
in the asymmetry \cite{ac93}).

In conclusion, we have considered in detail the integrated $CP$
violating asymmetries arising from heavy particle mixing, and studied
the effects that appear when the mass splittings are of the order of
the particle widths. The large enhancements which can be achieved can
be helpful to explain the observed baryon asymmetry of the Universe,
as we have exemplified with the study of a scenario for leptogenesis.

\vskip1cm

We would like to thank Francesco Vissani for
useful comments on the manuscript and for discussions. We also thank
A. Masiero for discussions.

\vfill\eject

\pagebreak

{\bf\large Figure captions:}

\bigskip

\noindent Figure 1: Diagrams 
which interfere to produce the $CP$ violation in the
heavy particle decay. Fig.~1b gives the so called vertex contribution
while Fig.~1c gives the  wave function one.

\bigskip

\noindent Figure 2: Wave function contribution to the $CP$ asymmetry,
 normalised to the vertex one, as a function of
$M_2/M_1$,  assuming $M^2_{12}=0$ and taking 
 $\Gamma^2_{22}=M^2_1/10$,
$\Gamma^2_{11}=|\Gamma^2_{12}|=\Gamma^2_{22}/10$.

\vfill\eject

\begin{figure}
\centerline{\epsfbox{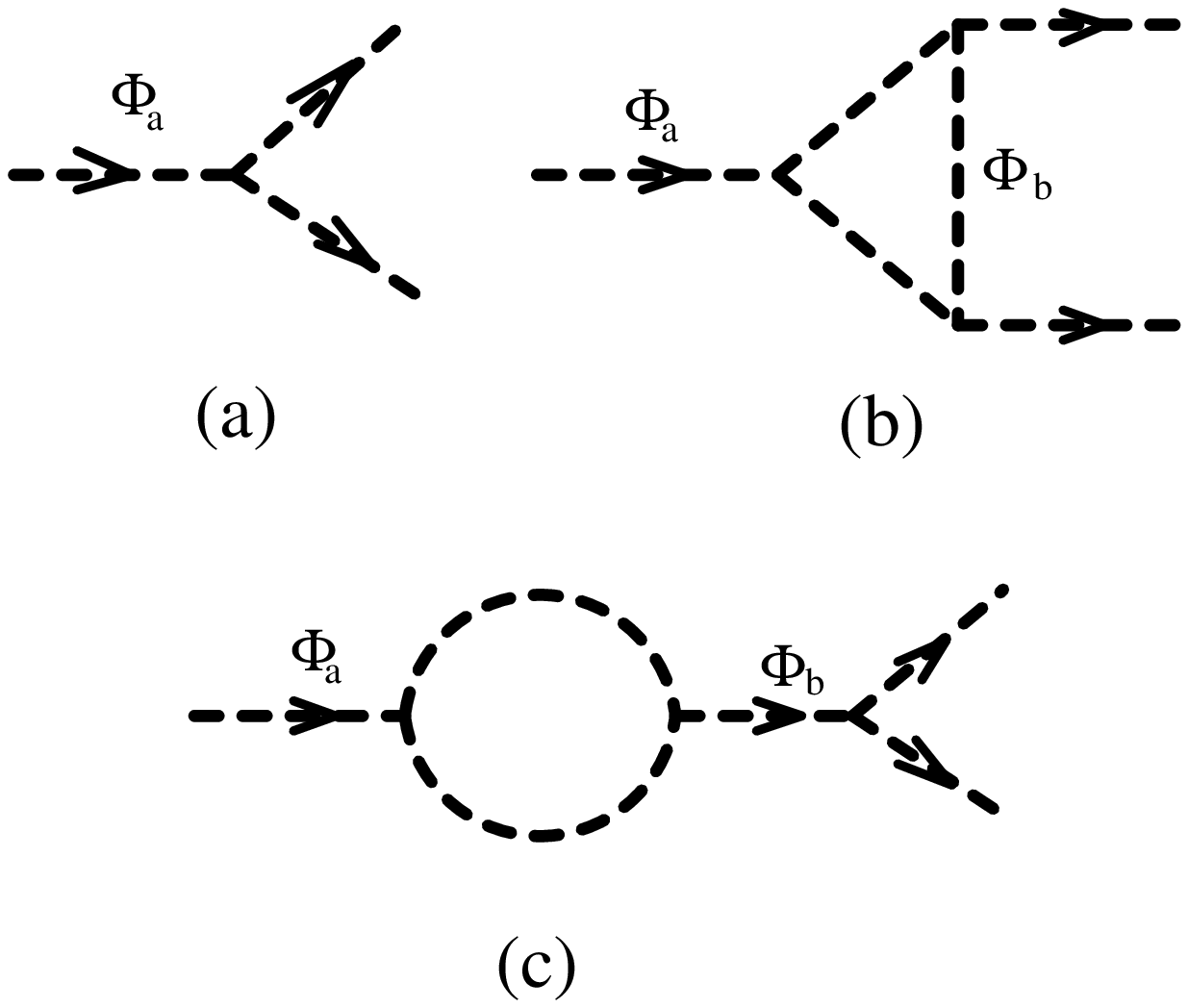}}
\end{figure}

\begin{figure}
\begin{center}
% GNUPLOT: LaTeX picture
\setlength{\unitlength}{0.240900pt}
\ifx\plotpoint\undefined\newsavebox{\plotpoint}\fi
\sbox{\plotpoint}{\rule[-0.500pt]{1.000pt}{1.000pt}}%
\begin{picture}(1500,900)(0,0)
\font\gnuplot=cmr10 at 10pt
\gnuplot
\sbox{\plotpoint}{\rule[-0.500pt]{1.000pt}{1.000pt}}%
\put(220.0,149.0){\rule[-0.500pt]{292.934pt}{1.000pt}}
\put(220.0,149.0){\rule[-0.500pt]{4.818pt}{1.000pt}}
\put(198,149){\makebox(0,0)[r]{0}}
\put(1416.0,149.0){\rule[-0.500pt]{4.818pt}{1.000pt}}
\put(220.0,331.0){\rule[-0.500pt]{4.818pt}{1.000pt}}
\put(198,331){\makebox(0,0)[r]{5}}
\put(1416.0,331.0){\rule[-0.500pt]{4.818pt}{1.000pt}}
\put(220.0,513.0){\rule[-0.500pt]{4.818pt}{1.000pt}}
\put(198,513){\makebox(0,0)[r]{10}}
\put(1416.0,513.0){\rule[-0.500pt]{4.818pt}{1.000pt}}
\put(220.0,695.0){\rule[-0.500pt]{4.818pt}{1.000pt}}
\put(198,695){\makebox(0,0)[r]{15}}
\put(1416.0,695.0){\rule[-0.500pt]{4.818pt}{1.000pt}}
\put(220.0,877.0){\rule[-0.500pt]{4.818pt}{1.000pt}}
\put(198,877){\makebox(0,0)[r]{20}}
\put(1416.0,877.0){\rule[-0.500pt]{4.818pt}{1.000pt}}
\put(220.0,113.0){\rule[-0.500pt]{1.000pt}{4.818pt}}
\put(220,68){\makebox(0,0){1}}
\put(220.0,857.0){\rule[-0.500pt]{1.000pt}{4.818pt}}
\put(423.0,113.0){\rule[-0.500pt]{1.000pt}{4.818pt}}
\put(423,68){\makebox(0,0){1.1}}
\put(423.0,857.0){\rule[-0.500pt]{1.000pt}{4.818pt}}
\put(625.0,113.0){\rule[-0.500pt]{1.000pt}{4.818pt}}
\put(625,68){\makebox(0,0){1.2}}
\put(625.0,857.0){\rule[-0.500pt]{1.000pt}{4.818pt}}
\put(828.0,113.0){\rule[-0.500pt]{1.000pt}{4.818pt}}
\put(828,68){\makebox(0,0){1.3}}
\put(828.0,857.0){\rule[-0.500pt]{1.000pt}{4.818pt}}
\put(1031.0,113.0){\rule[-0.500pt]{1.000pt}{4.818pt}}
\put(1031,68){\makebox(0,0){1.4}}
\put(1031.0,857.0){\rule[-0.500pt]{1.000pt}{4.818pt}}
\put(1233.0,113.0){\rule[-0.500pt]{1.000pt}{4.818pt}}
\put(1233,68){\makebox(0,0){1.5}}
\put(1233.0,857.0){\rule[-0.500pt]{1.000pt}{4.818pt}}
\put(1436.0,113.0){\rule[-0.500pt]{1.000pt}{4.818pt}}
\put(1436,68){\makebox(0,0){1.6}}
\put(1436.0,857.0){\rule[-0.500pt]{1.000pt}{4.818pt}}
\put(220.0,113.0){\rule[-0.500pt]{292.934pt}{1.000pt}}
\put(1436.0,113.0){\rule[-0.500pt]{1.000pt}{184.048pt}}
\put(220.0,877.0){\rule[-0.500pt]{292.934pt}{1.000pt}}
\put(45,495){\makebox(0,0){$\epsilon^w/ \epsilon^v$}}
\put(916,-22){\makebox(0,0){$M_2/M_1$}}
\put(220.0,113.0){\rule[-0.500pt]{1.000pt}{184.048pt}}
\put(220,145){\usebox{\plotpoint}}
\multiput(221.83,145.00)(0.495,4.798){32}{\rule{0.119pt}{9.700pt}}
\multiput(217.92,145.00)(20.000,168.867){2}{\rule{1.000pt}{4.850pt}}
\multiput(241.83,334.00)(0.496,3.859){34}{\rule{0.119pt}{7.869pt}}
\multiput(237.92,334.00)(21.000,143.667){2}{\rule{1.000pt}{3.935pt}}
\multiput(262.83,494.00)(0.495,2.753){32}{\rule{0.119pt}{5.700pt}}
\multiput(258.92,494.00)(20.000,97.169){2}{\rule{1.000pt}{2.850pt}}
\multiput(282.83,603.00)(0.495,1.500){32}{\rule{0.119pt}{3.250pt}}
\multiput(278.92,603.00)(20.000,53.254){2}{\rule{1.000pt}{1.625pt}}
\multiput(302.83,663.00)(0.495,0.580){32}{\rule{0.119pt}{1.450pt}}
\multiput(298.92,663.00)(20.000,20.990){2}{\rule{1.000pt}{0.725pt}}
\multiput(342.00,684.68)(0.740,-0.492){18}{\rule{1.788pt}{0.118pt}}
\multiput(342.00,684.92)(16.288,-13.000){2}{\rule{0.894pt}{1.000pt}}
\multiput(362.00,671.68)(0.503,-0.495){30}{\rule{1.303pt}{0.119pt}}
\multiput(362.00,671.92)(17.296,-19.000){2}{\rule{0.651pt}{1.000pt}}
\multiput(383.83,649.19)(0.495,-0.554){32}{\rule{0.119pt}{1.400pt}}
\multiput(379.92,652.09)(20.000,-20.094){2}{\rule{1.000pt}{0.700pt}}
\multiput(403.83,626.61)(0.496,-0.503){34}{\rule{0.119pt}{1.298pt}}
\multiput(399.92,629.31)(21.000,-19.307){2}{\rule{1.000pt}{0.649pt}}
\multiput(424.83,604.19)(0.495,-0.554){32}{\rule{0.119pt}{1.400pt}}
\multiput(420.92,607.09)(20.000,-20.094){2}{\rule{1.000pt}{0.700pt}}
\multiput(444.83,581.60)(0.495,-0.503){32}{\rule{0.119pt}{1.300pt}}
\multiput(440.92,584.30)(20.000,-18.302){2}{\rule{1.000pt}{0.650pt}}
\multiput(463.00,563.68)(0.478,-0.495){32}{\rule{1.250pt}{0.119pt}}
\multiput(463.00,563.92)(17.406,-20.000){2}{\rule{0.625pt}{1.000pt}}
\multiput(483.00,543.68)(0.560,-0.495){28}{\rule{1.417pt}{0.119pt}}
\multiput(483.00,543.92)(18.060,-18.000){2}{\rule{0.708pt}{1.000pt}}
\multiput(504.00,525.68)(0.599,-0.494){24}{\rule{1.500pt}{0.119pt}}
\multiput(504.00,525.92)(16.887,-16.000){2}{\rule{0.750pt}{1.000pt}}
\multiput(524.00,509.68)(0.599,-0.494){24}{\rule{1.500pt}{0.119pt}}
\multiput(524.00,509.92)(16.887,-16.000){2}{\rule{0.750pt}{1.000pt}}
\multiput(544.00,493.68)(0.723,-0.492){20}{\rule{1.750pt}{0.119pt}}
\multiput(544.00,493.92)(17.368,-14.000){2}{\rule{0.875pt}{1.000pt}}
\multiput(565.00,479.68)(0.740,-0.492){18}{\rule{1.788pt}{0.118pt}}
\multiput(565.00,479.92)(16.288,-13.000){2}{\rule{0.894pt}{1.000pt}}
\multiput(585.00,466.68)(0.803,-0.491){16}{\rule{1.917pt}{0.118pt}}
\multiput(585.00,466.92)(16.022,-12.000){2}{\rule{0.958pt}{1.000pt}}
\multiput(605.00,454.68)(0.878,-0.489){14}{\rule{2.068pt}{0.118pt}}
\multiput(605.00,454.92)(15.707,-11.000){2}{\rule{1.034pt}{1.000pt}}
\multiput(625.00,443.68)(1.022,-0.487){12}{\rule{2.350pt}{0.117pt}}
\multiput(625.00,443.92)(16.122,-10.000){2}{\rule{1.175pt}{1.000pt}}
\multiput(646.00,433.68)(0.969,-0.487){12}{\rule{2.250pt}{0.117pt}}
\multiput(646.00,433.92)(15.330,-10.000){2}{\rule{1.125pt}{1.000pt}}
\multiput(666.00,423.68)(1.226,-0.481){8}{\rule{2.750pt}{0.116pt}}
\multiput(666.00,423.92)(14.292,-8.000){2}{\rule{1.375pt}{1.000pt}}
\multiput(686.00,415.68)(1.082,-0.485){10}{\rule{2.472pt}{0.117pt}}
\multiput(686.00,415.92)(14.869,-9.000){2}{\rule{1.236pt}{1.000pt}}
\multiput(706.00,406.69)(1.502,-0.475){6}{\rule{3.250pt}{0.114pt}}
\multiput(706.00,406.92)(14.254,-7.000){2}{\rule{1.625pt}{1.000pt}}
\multiput(727.00,399.69)(1.420,-0.475){6}{\rule{3.107pt}{0.114pt}}
\multiput(727.00,399.92)(13.551,-7.000){2}{\rule{1.554pt}{1.000pt}}
\multiput(747.00,392.69)(1.420,-0.475){6}{\rule{3.107pt}{0.114pt}}
\multiput(747.00,392.92)(13.551,-7.000){2}{\rule{1.554pt}{1.000pt}}
\multiput(767.00,385.69)(1.708,-0.462){4}{\rule{3.583pt}{0.111pt}}
\multiput(767.00,385.92)(12.563,-6.000){2}{\rule{1.792pt}{1.000pt}}
\multiput(787.00,379.69)(1.811,-0.462){4}{\rule{3.750pt}{0.111pt}}
\multiput(787.00,379.92)(13.217,-6.000){2}{\rule{1.875pt}{1.000pt}}
\multiput(808.00,373.71)(2.358,-0.424){2}{\rule{4.250pt}{0.102pt}}
\multiput(808.00,373.92)(11.179,-5.000){2}{\rule{2.125pt}{1.000pt}}
\multiput(828.00,368.71)(2.358,-0.424){2}{\rule{4.250pt}{0.102pt}}
\multiput(828.00,368.92)(11.179,-5.000){2}{\rule{2.125pt}{1.000pt}}
\multiput(848.00,363.71)(2.528,-0.424){2}{\rule{4.450pt}{0.102pt}}
\multiput(848.00,363.92)(11.764,-5.000){2}{\rule{2.225pt}{1.000pt}}
\put(869,356.92){\rule{4.818pt}{1.000pt}}
\multiput(869.00,358.92)(10.000,-4.000){2}{\rule{2.409pt}{1.000pt}}
\multiput(889.00,354.71)(2.358,-0.424){2}{\rule{4.250pt}{0.102pt}}
\multiput(889.00,354.92)(11.179,-5.000){2}{\rule{2.125pt}{1.000pt}}
\put(909,347.92){\rule{4.818pt}{1.000pt}}
\multiput(909.00,349.92)(10.000,-4.000){2}{\rule{2.409pt}{1.000pt}}
\put(929,343.92){\rule{5.059pt}{1.000pt}}
\multiput(929.00,345.92)(10.500,-4.000){2}{\rule{2.529pt}{1.000pt}}
\put(950,340.42){\rule{4.818pt}{1.000pt}}
\multiput(950.00,341.92)(10.000,-3.000){2}{\rule{2.409pt}{1.000pt}}
\put(970,336.92){\rule{4.818pt}{1.000pt}}
\multiput(970.00,338.92)(10.000,-4.000){2}{\rule{2.409pt}{1.000pt}}
\put(990,333.42){\rule{4.818pt}{1.000pt}}
\multiput(990.00,334.92)(10.000,-3.000){2}{\rule{2.409pt}{1.000pt}}
\put(1010,330.42){\rule{5.059pt}{1.000pt}}
\multiput(1010.00,331.92)(10.500,-3.000){2}{\rule{2.529pt}{1.000pt}}
\put(1031,327.42){\rule{4.818pt}{1.000pt}}
\multiput(1031.00,328.92)(10.000,-3.000){2}{\rule{2.409pt}{1.000pt}}
\put(1051,324.42){\rule{4.818pt}{1.000pt}}
\multiput(1051.00,325.92)(10.000,-3.000){2}{\rule{2.409pt}{1.000pt}}
\put(1071,321.42){\rule{4.818pt}{1.000pt}}
\multiput(1071.00,322.92)(10.000,-3.000){2}{\rule{2.409pt}{1.000pt}}
\put(1091,318.92){\rule{5.059pt}{1.000pt}}
\multiput(1091.00,319.92)(10.500,-2.000){2}{\rule{2.529pt}{1.000pt}}
\put(1112,316.42){\rule{4.818pt}{1.000pt}}
\multiput(1112.00,317.92)(10.000,-3.000){2}{\rule{2.409pt}{1.000pt}}
\put(1132,313.92){\rule{4.818pt}{1.000pt}}
\multiput(1132.00,314.92)(10.000,-2.000){2}{\rule{2.409pt}{1.000pt}}
\put(1152,311.92){\rule{5.059pt}{1.000pt}}
\multiput(1152.00,312.92)(10.500,-2.000){2}{\rule{2.529pt}{1.000pt}}
\put(1173,309.92){\rule{4.818pt}{1.000pt}}
\multiput(1173.00,310.92)(10.000,-2.000){2}{\rule{2.409pt}{1.000pt}}
\put(1193,307.92){\rule{4.818pt}{1.000pt}}
\multiput(1193.00,308.92)(10.000,-2.000){2}{\rule{2.409pt}{1.000pt}}
\put(1213,305.92){\rule{4.818pt}{1.000pt}}
\multiput(1213.00,306.92)(10.000,-2.000){2}{\rule{2.409pt}{1.000pt}}
\put(1233,303.92){\rule{5.059pt}{1.000pt}}
\multiput(1233.00,304.92)(10.500,-2.000){2}{\rule{2.529pt}{1.000pt}}
\put(1254,301.92){\rule{4.818pt}{1.000pt}}
\multiput(1254.00,302.92)(10.000,-2.000){2}{\rule{2.409pt}{1.000pt}}
\put(1274,299.92){\rule{4.818pt}{1.000pt}}
\multiput(1274.00,300.92)(10.000,-2.000){2}{\rule{2.409pt}{1.000pt}}
\put(1294,297.92){\rule{4.818pt}{1.000pt}}
\multiput(1294.00,298.92)(10.000,-2.000){2}{\rule{2.409pt}{1.000pt}}
\put(1314,296.42){\rule{5.059pt}{1.000pt}}
\multiput(1314.00,296.92)(10.500,-1.000){2}{\rule{2.529pt}{1.000pt}}
\put(1335,294.92){\rule{4.818pt}{1.000pt}}
\multiput(1335.00,295.92)(10.000,-2.000){2}{\rule{2.409pt}{1.000pt}}
\put(1355,292.92){\rule{4.818pt}{1.000pt}}
\multiput(1355.00,293.92)(10.000,-2.000){2}{\rule{2.409pt}{1.000pt}}
\put(1375,291.42){\rule{4.818pt}{1.000pt}}
\multiput(1375.00,291.92)(10.000,-1.000){2}{\rule{2.409pt}{1.000pt}}
\put(1395,290.42){\rule{5.059pt}{1.000pt}}
\multiput(1395.00,290.92)(10.500,-1.000){2}{\rule{2.529pt}{1.000pt}}
\put(1416,288.92){\rule{4.818pt}{1.000pt}}
\multiput(1416.00,289.92)(10.000,-2.000){2}{\rule{2.409pt}{1.000pt}}
\put(321.0,687.0){\rule[-0.500pt]{5.059pt}{1.000pt}}
\put(1436,290){\usebox{\plotpoint}}
\sbox{\plotpoint}{\rule[-0.250pt]{0.500pt}{0.500pt}}%
\multiput(378,877)(2.330,-12.233){2}{\usebox{\plotpoint}}
\multiput(382,856)(3.291,-12.011){6}{\usebox{\plotpoint}}
\multiput(402,783)(4.240,-11.709){5}{\usebox{\plotpoint}}
\multiput(423,725)(4.790,-11.495){4}{\usebox{\plotpoint}}
\multiput(443,677)(5.569,-11.139){4}{\usebox{\plotpoint}}
\multiput(463,637)(6.455,-10.650){3}{\usebox{\plotpoint}}
\multiput(483,604)(7.304,-10.086){3}{\usebox{\plotpoint}}
\multiput(504,575)(7.780,-9.724){3}{\usebox{\plotpoint}}
\multiput(524,550)(8.377,-9.215){2}{\usebox{\plotpoint}}
\multiput(544,528)(9.235,-8.355){2}{\usebox{\plotpoint}}
\multiput(565,509)(9.489,-8.065){2}{\usebox{\plotpoint}}
\multiput(585,492)(9.963,-7.472){2}{\usebox{\plotpoint}}
\multiput(605,477)(10.202,-7.141){2}{\usebox{\plotpoint}}
\multiput(625,463)(10.812,-6.179){2}{\usebox{\plotpoint}}
\multiput(646,451)(10.679,-6.407){2}{\usebox{\plotpoint}}
\multiput(666,439)(11.139,-5.569){2}{\usebox{\plotpoint}}
\multiput(686,429)(11.356,-5.110){2}{\usebox{\plotpoint}}
\multiput(706,420)(11.446,-4.906){2}{\usebox{\plotpoint}}
\put(738.53,406.39){\usebox{\plotpoint}}
\multiput(747,403)(11.563,-4.625){2}{\usebox{\plotpoint}}
\multiput(767,395)(11.928,-3.578){2}{\usebox{\plotpoint}}
\put(797.17,385.61){\usebox{\plotpoint}}
\multiput(808,382)(11.928,-3.578){2}{\usebox{\plotpoint}}
\multiput(828,376)(12.081,-3.020){2}{\usebox{\plotpoint}}
\multiput(848,371)(11.974,-3.421){2}{\usebox{\plotpoint}}
\put(881.18,362.56){\usebox{\plotpoint}}
\multiput(889,361)(12.081,-3.020){2}{\usebox{\plotpoint}}
\put(917.60,354.28){\usebox{\plotpoint}}
\multiput(929,352)(12.233,-2.330){2}{\usebox{\plotpoint}}
\multiput(950,348)(12.211,-2.442){2}{\usebox{\plotpoint}}
\put(978.70,342.26){\usebox{\plotpoint}}
\multiput(990,340)(12.316,-1.847){2}{\usebox{\plotpoint}}
\multiput(1010,337)(12.233,-2.330){2}{\usebox{\plotpoint}}
\put(1040.04,331.64){\usebox{\plotpoint}}
\multiput(1051,330)(12.316,-1.847){2}{\usebox{\plotpoint}}
\multiput(1071,327)(12.316,-1.847){2}{\usebox{\plotpoint}}
\put(1101.69,322.98){\usebox{\plotpoint}}
\multiput(1112,322)(12.316,-1.847){2}{\usebox{\plotpoint}}
\multiput(1132,319)(12.391,-1.239){2}{\usebox{\plotpoint}}
\put(1163.47,315.36){\usebox{\plotpoint}}
\multiput(1173,314)(12.391,-1.239){2}{\usebox{\plotpoint}}
\multiput(1193,312)(12.391,-1.239){2}{\usebox{\plotpoint}}
\put(1225.37,308.76){\usebox{\plotpoint}}
\multiput(1233,308)(12.397,-1.181){2}{\usebox{\plotpoint}}
\put(1262.56,305.14){\usebox{\plotpoint}}
\multiput(1274,304)(12.391,-1.239){2}{\usebox{\plotpoint}}
\multiput(1294,302)(12.391,-1.239){2}{\usebox{\plotpoint}}
\put(1324.52,299.00){\usebox{\plotpoint}}
\multiput(1335,298)(12.438,-0.622){2}{\usebox{\plotpoint}}
\multiput(1355,297)(12.391,-1.239){2}{\usebox{\plotpoint}}
\put(1386.60,294.42){\usebox{\plotpoint}}
\multiput(1395,294)(12.397,-1.181){2}{\usebox{\plotpoint}}
\put(1423.84,291.61){\usebox{\plotpoint}}
\put(1436,291){\usebox{\plotpoint}}
\end{picture}

\end{center}
\end{figure}
\vfill

\end{document}